\documentclass{JHEP3}

\usepackage{epsfig}
\usepackage{cite}
  \usepackage{graphicx}
  \usepackage{amssymb}

\preprint{DCP-09-01\\
}

\title{A method of diagonalization for sfermion mass matrices}

\author{Alfredo Aranda$\rm \, ^{a,b}$\footnote{Email: {\tt
fefo@ucol.mx}}, J.L. D\'iaz-Cruz$\rm \, ^{a,c}$\footnote{Email: {\tt
ldiaz@sirio.ifuap.buap.mx}}, R. Noriega-Papaqui$\rm \, ^{a,d}$\footnote{Email: {\tt
rnoriega@uaeh.edu.mx}}\\

{\it $\rm ^a$ Dual CP Institute of High Energy Physics, M\'exico}\\
{\it $\rm ^b$ Facultad de Ciencias, CUICBAS, Universidad de Colima,
  Bernal D\'{i}az del Castillo 340, Colima, Colima, M\'exico} \\
{\it $\rm ^c$FCFM, BUAP, Av. San Claudio y 18 Sur, C. P.
72570, Puebla, Pue., M\'exico} \\
{\it $\rm ^d$ Centro de Investigaci\'on en Matem\'aticas, UAEH,
Carr. Pachuca-Tulancingo Km. 4.5, C.P. 42184, Pachuca, Hgo., M\'exico}}

\abstract{We present a method of diagonalization for the sfermion mass matrices of the
minimal supersymmetric standard model (MSSM). It
provides analytical expressions for the masses and mixing angles of rather general
hermitian sfermion mass matrices, and allows the study of scenarios that extend the
usual constrained - MSSM. Three signature cases are presented
explicitly and a general study of flavor changing neutral current processes is 
outlined in the discussion.}


\begin{document}

\newcommand{\bfig}{\begin{figure}}
\newcommand{\efig}{\end{figure}}
\newcommand{\be}{\begin{equation}}
\newcommand{\ee}{\end{equation}}
\newcommand{\bea}{\begin{eqnarray}}
\newcommand{\eea}{\end{eqnarray}}
\newcommand{\bc}{\begin{center}}
\newcommand{\ec}{\end{center}}
\newcommand{\nn}{\nonumber}

\section{Introduction}\label{intro}

The Minimal Supersymmetric Standard Model (MSSM) is one of the main extensions of the Standard Model (SM). 
Its main motivation is its natural resolution to the hierarchy problem. Its basic 
structure is obtained by assuming that there is (low-scale) supersymmetry which then
immediately predicts the existence of a so-called super-partner for each of the SM particles. Additionally it requires
an extension of the usual (minimal) scalar sector of the SM to two Higgs doublets~\cite{Haber:1984rc,Aitchison:2005cf}. 

Supersymmetry, if at all present, must be broken and the MSSM must contain this information\cite{Chung:2003fi}. Since the actual
mechanism of supersymmetry breaking that would lead to the MSSM is still an open problem, the best one can do is
to assume that supersymmetry is broken softly (in order not to spoil the supersymmetry based solution to the hierarchy 
problem). The result is that all super-partners are assumed to have soft masses that in turn become
unknown free parameters of the model. One can then try to use physically sensible assumptions as to their magnitudes and
hierarchies and perform phenomenological studies that eventually lead to a constrained parameter space.

Of particular interest are the sfermions mass matrices. These matrices receive contributions from 
the so-called trilinear terms, also called A-terms, which can be strong sources of flavor changing neutral currents 
(FCNC)~\cite{Misiak:1997ei}. 
The sfermions mass matrices are $6\times 6$ matrices usually written in the basis (for the up-type squarks, for example) 
$\{ \tilde{u}_L , \  \tilde{c}_L , \ \tilde{t}_L , \ \tilde{u}_R , \ \tilde{c}_R , \ \tilde{u}_R , \}$. The entries in these
matrices are arbitrary and will be determined by the supersymmetry breaking mechanism that leads to the MSSM. 

One can start reducing the arbitrariness in the parameter space by assuming, for example, that the matrices are hermitian, or by  
imposing certain relations among the entries through the use of flavor symmetries. Also, considering some specific scenarios
of supersymmetry breaking (gauge or gravity mediated) one can impose degeneracies (universality) and or hierarchies among the 
different  parameters~\cite{Kobayashi:2000br}. 
In practice, however, most of the phenomenological studies have been done under the physically sound assumption 
that only the third generation sfermions contribute significantly to the A-terms. Besides being somehow natural (due to the contribution
of the Yukawa couplings to the A-terms), this assumption is also convenient if one is interested in obtaining analytical expressions.
The diagonalization of the sfermions mass matrices in this case can be easily done while more general cases are
done numerically. Other scenarios exist where some of the universality is broken with some specific non-zero entries in the 
A-terms~\cite{DiazCruz:2001gf,GomezBock:2008hz,Hiller:2008sv}.

Motivated by this situation we present a method of diagonalization for the $6\times 6$ matrices that gives analytical 
expressions and that can be used under more general cases than those usually studied. In Section~\ref{method} we present
the method in detail and consider three different cases. We explicitly express the physical sfermion (squared) masses and
give expressions for the mixing angles. This is followed by a discussion in Section~\ref{discussion} where some comments
are given as to the relevance that this method can have in model-independent analysis of certain FCNC processes. Finally 
we present our conclusions.

\section{The method}\label{method}

Consider the up-type squarks ($6 \times 6$) mass matrix
\begin{eqnarray}\label{mu}
  \tilde{M}_u^2 = 
  \left(
  \begin{array}{cc}
    M_{LL}^2 & M_{LR}^2 \\
    M_{LR}^{2 \dagger} & M_{RR}^2
  \end{array}
  \right) \ ,
\end{eqnarray}
where $ M_{LL}^2$, $M_{RR}^2$, and $M_{LR}^2$ are $3 \times 3$ matrices given by
\begin{eqnarray}
  \label{mll}
  M_{LL}^2 & = & M_{\tilde{Q}}^2 + M_u^2 + \frac{1}{6}\cos 2\beta (4m_W^2-m_Z^2) \ , \\
  \label{mrr}
  M_{RR}^2 & = & M_{\tilde{u}}^2 + M_u^2 + \frac{2}{3}\cos 2\beta \sin^2\theta_w m_Z^2 \ , \\
  \label{mlr}
  M_{LR}^2 & = & A_u \frac{v}{\sqrt{2}}\sin\beta - M_u^2 \mu \cot\beta \ ,
\end{eqnarray}
where $v^2=v_u^2+v_d^2$, with $v_{u(d)} = \langle H_{u(d)} \rangle$ and $\tan\beta = v_u/v_d$. Note
that these matrices receive contributions from soft-breaking terms ($M_{\tilde{Q}}^2$, $M_{\tilde{u}}^2$, 
$A_u$, and $\mu$), from the breaking of electroweak symmetry ($M_u^2$), and from the D-terms in the 
lagrangian.

The contributions involving soft-breaking terms constitute free parameters and there is no precise
way to fix them. As a result, the mass matrix Eq.~(\ref{mu}) is completely arbitrary. Furthermore, the
so-called A-terms can contribute to flavor changing neutral current (FCNC) processes and must be handled
with care. From now on we work under the assumption that $\tilde{M}_u^2$ is a hermitian matrix

Given these considerations we proceed to the diagonalization of the mass matrix $\tilde{M}_u^2$
by constructing a matrix $V_{\tilde{u}}$ such that
\begin{eqnarray}\label{vmv}
\tilde{M}_{uD}^2 = V_{\tilde{u}}^{\dagger} \tilde{M}_u^2 V_{\tilde{u}} \ .
\end{eqnarray}

We reparametrize the mass matrix Eq.~(\ref{mu}) as
\begin{eqnarray}\label{muabc}
  \tilde{M}_u^2 = 
  \left(
  \begin{array}{cc}
    A & B \\
    B^{\dagger} & C
  \end{array}
  \right) \ ,
\end{eqnarray}
and introduce a unitary ($6\times 6$) matrix $U$ of the form
\begin{eqnarray}\label{u}
  U = 
  \left(
  \begin{array}{cc}
    U_L & 0 \\
    0 & U_R
  \end{array}
  \right) \ ,
\end{eqnarray}
with $U_{L(R)}$ unitary $3\times 3$ matrices. Then, consider the following expression:
\begin{eqnarray} \label{mstar}
  \tilde{M}^2_{u*}=U^{\dagger} \tilde{M}_u^2 U =   
  \left(
  \begin{array}{cc}
    U_L^{\dagger} A U_L & U_L^{\dagger} B U_R \\
    (U_L^{\dagger} B U_R)^{\dagger} & U_R^{\dagger} C U_R
  \end{array}
  \right) \ .
\end{eqnarray}

Since $B$ is hermitian (by assumption), it is clear that we can use the matrices $U_L$ and $U_R$
to diagonalize the matrix $B$ (with eigenvalues $b_1$, $b_2$, and $b_3$). 
On the other hand, the matrices $A$ and $C$ will not in general
be diagonalized by these matrices, see Appendix~\ref{appendix}.
Let us continue by assuming for the moment that the matrices $A$ and $B$ satisfy
this condition (we present explicit examples below),
we then obtain a matrix with form
\begin{eqnarray} \label{form}
  \tilde{M}^2_{u*} = U^{\dagger} \tilde{M}_u^2 U =   
  \left(
  \begin{array}{ccc|ccc}
    a_1 & 0 & 0 & b_1 & 0 & 0 \\
    0 & a_2 & 0 & 0 & b_2 & 0 \\
    0 & 0 & a_3 & 0 & 0 & b_3 \\
    \hline
    b_1 & 0 & 0 & c_1 & 0 & 0 \\
    0 & b_2 & 0 & 0 & c_2 & 0 \\
    0 & 0 & b_3 & 0 & 0 & c_3 
  \end{array}
  \right) \ ,
\end{eqnarray}
where $a_1, \ a_2, \ a_3 $ are the eigenvalues of $A$ and similarly $c_1, \ c_2, \ c_3$ the ones for $C$.

Going back to the original mass matrix $\tilde{M}_u^2$ we note that it is written in the
basis $\hat{u}_{LLLRRR} \equiv $ 
\{$\tilde{u}_L$, $\tilde{c}_L$, $\tilde{t}_L$, $\tilde{u}_R$, $\tilde{c}_R$, $\tilde{t}_R$\}. The next
step in the procedure is to express the mass matrix in the different basis 
$\hat{u}_{LRLRLR} \equiv $ 
\{$\tilde{u}_L$, $\tilde{u}_R$, $\tilde{c}_L$, $\tilde{c}_R$, $\tilde{t}_L$, $\tilde{t}_R$\}. This can be
easily accomplished using the matrix $T$ defined by
\begin{eqnarray}\label{t}
  T = 
  \left(
  \begin{array}{cc|cc|cc}
    1 & 0 & 0 & 0 & 0 & 0 \\
    0 & 0 & 0 & 1 & 0 & 0 \\
    \hline
    0 & 1 & 0 & 0 & 0 & 0 \\
    0 & 0 & 0 & 0 & 1 & 0 \\
    \hline
    0 & 0 & 1 & 0 & 0 & 0 \\
    0 & 0 & 0 & 0 & 0 & 1 
  \end{array}
  \right) \ ,
\end{eqnarray}
and we obtain the matrix given by
\begin{eqnarray}\label{mubd}
  \tilde{M}^2_{uBD} = T  \tilde{M}^2_{u*} T^{\dagger} =
  \left(
  \begin{array}{cc|cc|cc}
    a_1 & b_1 & 0 & 0 & 0 & 0 \\
    b_1 & c_1 & 0 & 0 & 0 & 0 \\
    \hline
    0 & 0 & a_2 & b_2 & 0 & 0 \\
    0 & 0 & b_2 & c_2 & 0 & 0 \\
    \hline
    0 & 0 & 0 & 0 & a_3 & b_3 \\
    0 & 0 & 0 & 0 & b_3 & c_3 
  \end{array}
  \right) 
  =
  \left(
  \begin{array}{c|c|c}
    G_{1} &  0 & 0 \\
    \hline
    0 & G_{2} & 0 \\
    \hline
    0 & 0 & G_{3}
  \end{array}
  \right) \ .
\end{eqnarray}

Note that since all the $3 \times 3$ blocks in
the matrix in Eq.~(\ref{mstar}) are hermitian matrices, then their eigenvalues are real numbers (see Eq.~(\ref{form})). 
Furthermore, since the off-diagonal blocks are the same ($B^{\dagger}=B$), the off-diagonal entries in each of the
$G_{x}$ ($x = 1, \ 2, \ 3$) are equal and thus the $G_{x}$ are $2 \times 2$ 
real symmetric matrices. As such they can be
diagonalized by orthogonal matrices $R_{x}$ in the following way:
\begin{eqnarray} \label{rmubdr}
G_{1D} = R_{1}^{\dagger} G_{1}  R_{1} \ , \qquad
G_{2D} = R_{2}^{\dagger} G_{2} R_{2} \ , \qquad
G_{3D} = R_{3}^{\dagger} G_{3}  R_{3} \ ,
\end{eqnarray}  
with eigenvalues of $G_{x}$ denoted by $\lambda_{x,1}$ and  $\lambda_{x,2}$ and given by
\begin{eqnarray} \label{eigenvalues}
\lambda_{x,1} = \frac{1}{2} \left( a_x + c_x - \Delta_x \right), \qquad
\lambda_{x,2} = \frac{1}{2} \left( a_x + c_x + \Delta_x \right),
\end{eqnarray}
with $\Delta_x = \sqrt{(a_x-c_x)^2+ 4 b_x^2}$, 
and where the matrices $R_{x}$ can be parametrized in terms of mixing angles $\theta_{x}$ in the usual way:
\begin{eqnarray}\label{rtheta}
  R_{x} = 
  \left(
  \begin{array}{cc}
    \cos\theta_{x} & \sin\theta_{x} \\
    -\sin\theta_{x} & \cos\theta_{x} 
  \end{array}
  \right) \ ,
\end{eqnarray}
with 
\begin{eqnarray} \label{angles}
\cos\theta_x = \frac{1}{\sqrt{2}}\left(1-\frac{a_x-c_x}{\Delta_x}\right)^{1/2}, \qquad 
\sin\theta_x = \frac{1}{\sqrt{2}}\left(1+\frac{a_x-c_x}{\Delta_x}\right)^{1/2} \ .
\end{eqnarray}

Finally, defining the $6 \times 6$ matrix ${\cal R}_{\tilde{u}}= {\rm diag}(R_{1}, \ R_{2}, \ R_{3})$, 
we express the
diagonal matrix $\tilde{M}_{uD}^2$ as
\begin{eqnarray} \label{mudfinal}
\tilde{M}_{uD}^2 = V_{\tilde{u}}^{\dagger} \tilde{M}_u^2 V_{\tilde{u}} = {\cal R}_{\tilde{u}}^{\dagger} T U^{\dagger} 
\tilde{M}_u^2 U T^{\dagger} {\cal R}_{\tilde{u}} \ ,
\end{eqnarray}
and thus the matrix $V_{\tilde{u}} = U T^{\dagger} {\cal R}_{\tilde{u}}$ diagonalizes the (up-type) squark 
mass matrix $\tilde{M}_u^2$. Note that the same procedure is easily extended to the down-type squark mass 
matrix as well as to the slepton mass matrix where one obtains (in obvious notation)
\begin{eqnarray} \label{mdfinalnlfinal}
\tilde{M}_{dD}^2 & = & V_{\tilde{d}}^{\dagger} \tilde{M}_d^2 V_{\tilde{d}} = {\cal R}_{\tilde{d}}^{\dagger} T D^{\dagger} 
\tilde{M}_d^2 D T^{\dagger} {\cal R}_{\tilde{d}} \ , \\
\tilde{M}_{lD}^2 & = & V_{\tilde{l}}^{\dagger} \tilde{M}_l^2 V_{\tilde{l}} = {\cal R}_{\tilde{l}}^{\dagger} T L^{\dagger} 
\tilde{M}_l^2 L T^{\dagger} {\cal R}_{\tilde{l}} \ .
\end{eqnarray}

As discussed after Eq.~(\ref{mstar}), these results apply only to those cases in which the matrices $M_{LL}^2$ 
and $M_{RR}^2$ are related 
to the matrix $M_{LR}^2$ in such a way as to be diagonalized once $M_{LR}^2$ is (see Appendix~\ref{appendix}).

We now proceed to show the application of this method to three different cases.

\subsection{Case a}\label{casea}

As a first example of the application of the method described above, we choose a scenario where both
$M_{LL}^2$ and $M_{RR}^2$ are proportional to the identity ${\cal I}_3$ and $M_{LR}^2$ is an arbitrary hermitian
matrix (we show the analysis for the up-type squark mass matrix):
\begin{eqnarray}\label{relationa}
M^2_{LL} = a {\cal I}_3, \qquad M^2_{RR} = c {\cal I}_3, \qquad M^2_{LR} = (M^2_{LR})^{\dagger} .
\end{eqnarray}
Then, the mass matrix becomes (see Eq.~(\ref{mu})
\begin{eqnarray}\label{mua}
  \tilde{M}_u^2 = 
  \left(
  \begin{array}{cc}
    a {\cal I}_3 & M_{LR}^2 \\
    M_{LR}^{2 \dagger} &  c {\cal I}_3
  \end{array}
  \right) \ ,
\end{eqnarray}
and using Eq.~(\ref{mstar}) we obtain
\begin{eqnarray}\label{mstara}
  \tilde{M}^2_{u*}=U^{\dagger} \tilde{M}_u^2 U =   
  \left(
  \begin{array}{cc}
    a \ U_L^{\dagger} {\cal I}_3 U_L & U_L^{\dagger} M^2_{LR} U_R \\
    (U_L^{\dagger} M_{LR}^2 U_R)^{\dagger} & c \ U_R^{\dagger} {\cal I}_3 U_R
  \end{array}
  \right) = 
  \left(
  \begin{array}{ccc|ccc}
    a & 0 & 0 & b_1 & 0 & 0 \\
    0 & a & 0 & 0 & b_2 & 0 \\
    0 & 0 & a & 0 & 0 & b_3 \\
    \hline
    b_1 & 0 & 0 & c & 0 & 0 \\
    0 & b_2 & 0 & 0 & c & 0 \\
    0 & 0 & b_3 & 0 & 0 & c 
  \end{array}
  \right) \ ,
\end{eqnarray}
where the unitary matrices $U_L$ and $U_R$ diagonalize $M_{LR}^2$ with eigenvalues
$b_x$. Since all matrices are hermitian we have $a$, $c$, $b_x \in \Re$.

The next step is to rotate this matrix to the basis $\hat{u}_{LRLRLR}$ using the matrix $T$ in Eq.~(\ref{t}):
\begin{eqnarray}\label{mubda}
  \tilde{M}^2_{uBD} = T  \tilde{M}^2_{u*} T^{\dagger} =
  \left(
  \begin{array}{cc|cc|cc}
    a & b_1 & 0 & 0 & 0 & 0 \\
    b_1 & c & 0 & 0 & 0 & 0 \\
    \hline
    0 & 0 & a & b_2 & 0 & 0 \\
    0 & 0 & b_2 & c & 0 & 0 \\
    \hline
    0 & 0 & 0 & 0 & a & b_3 \\
    0 & 0 & 0 & 0 & b_3 & c 
  \end{array}
  \right) 
  =
  \left(
  \begin{array}{c|c|c}
    G_{1} &  0 & 0 \\
    \hline
    0 & G_{2} & 0 \\
    \hline
    0 & 0 & G_{3}
  \end{array}
  \right) \ .
\end{eqnarray}

From Eq.~(\ref{eigenvalues}) we immediately obtain:
\begin{eqnarray} \label{eigenvaluesa}
  \lambda_{\tilde{U}1} =  \frac{1}{2} \left(a + c - \Delta_{\tilde{U}} \right), \qquad
  \lambda_{\tilde{U}2} =  \frac{1}{2} \left(a + c + \Delta_{\tilde{U}}\right) \ ,
\end{eqnarray}
where we have renamed $x = 1,\ 2,\ 3$ by $\tilde{U} = \tilde{u},\ \tilde{c},\ \tilde{t}$, and where
$\Delta_{\tilde{U}}=+\sqrt{(a-c)^2+4b_{\tilde{U}}^2}$. Similarly, for the mixing angles in $R_{\tilde{U}}$ we obtain
\begin{eqnarray} \label{anglesa}
  \sin\theta_{\tilde{U}} = \frac{1}{\sqrt{2}}\left( 1+\frac{a-c}{\Delta_{\tilde{U}}}\right)^{1/2},
  \qquad
  \cos\theta_{\tilde{U}} = \frac{1}{\sqrt{2}}\left( 1-\frac{a-c}{\Delta_{\tilde{U}}}\right)^{1/2} \ ,
\end{eqnarray}
and so we finally arrive at the desired up-type squark mass matrix
\begin{eqnarray}\label{finala}
  \tilde{M}^2_{uD} = V^{\dagger} \tilde{M}^2_u V = {\rm diag}
  \left(\lambda_{\tilde{u}1} \ \lambda_{\tilde{u}2} \ \lambda_{\tilde{c}1} \ \lambda_{\tilde{c}2} \ 
  \lambda_{\tilde{t}1} \ \lambda_{\tilde{t}2} \right) \ .
\end{eqnarray}

We can now make some comments regarding the spectrum. If we consider a hierarchy for the
$M_{LR}^2$ eigenvalues of the form $|b_{\tilde{u}}| < |b_{\tilde{c}}| < |b_{\tilde{t}}|$, then
this implies that $\Delta_{\tilde{u}} < \Delta_{\tilde{c}} < \Delta_{\tilde{t}}$ which then leads to
\begin{eqnarray}\label{spectruma}
  \lambda_{\tilde{t}1} < \lambda_{\tilde{c}1} < \lambda_{\tilde{u}1} <
  \lambda_{\tilde{u}2} < \lambda_{\tilde{c}2} < \lambda_{\tilde{t}2} \ .
\end{eqnarray}

We see that in this case the lightest u-type squark has squared mass $\lambda_{\tilde{t}1}$, and the
heaviest has a squared mass $\lambda_{\tilde{t}2}$~\footnote{These relations for the spectrum are valid 
at the supersymmetry breaking scale and can be affected by their running down to the electroweak scale.}.

\subsection{Case b}\label{caseb}
Another interesting scenario consists of having an arbitrary (hermitian)
$M_{LL}^2$ and both $M_{RR}^2$ and $M_{LR}^2$ proportional to the identity ${\cal I}_3$ matrix: 
(again, we show the analysis for the up-type squark mass matrix):
\begin{eqnarray}\label{relationb}
M^2_{LL} = (M^2_{LL})^{\dagger}, \qquad M^2_{RR} = c {\cal I}_3, \qquad M^2_{LR} = b {\cal I}_3  .
\end{eqnarray}
Then, the mass matrix becomes (see Eq.~(\ref{mu}))
\begin{eqnarray}\label{mub}
  \tilde{M}_u^2 = 
  \left(
  \begin{array}{cc}
    M_{LL}^2 & b {\cal I}_3 \\
    b {\cal I}_3 &  c {\cal I}_3
  \end{array}
  \right) \ ,
\end{eqnarray}
and using Eq.~(\ref{mstar}) we obtain
\begin{eqnarray}\label{mstarb}
  \tilde{M}^2_{u*}=U^{\dagger} \tilde{M}_u^2 U =   
  \left(
  \begin{array}{cc}
    U_L^{\dagger} M_{LL}^2 U_L & b \ U_L^{\dagger} {\cal I}_3 U_L \\
    (b \ U_L^{\dagger} {\cal I}_3 U_L)^{\dagger} & c \ U_L^{\dagger} {\cal I}_3 U_L
  \end{array}
  \right) = 
  \left(
  \begin{array}{ccc|ccc}
    a_1 & 0 & 0 & b & 0 & 0 \\
    0 & a_2 & 0 & 0 & b & 0 \\
    0 & 0 & a_3 & 0 & 0 & b \\
    \hline
    b & 0 & 0 & c & 0 & 0 \\
    0 & b & 0 & 0 & c & 0 \\
    0 & 0 & b & 0 & 0 & c 
  \end{array}
  \right) \ ,
\end{eqnarray}
where  $a_x$, $b$, $c \in \Re$ (since all matrices are hermitian). Note that in order to obtain diagonal matrices 
in the $1-2$ and $2-1$ sub-blocks we require (wlog) $U_R = U_L$.

Applying the method we obtain:
\begin{eqnarray} \label{eigenvaluesb}
  \lambda_{\tilde{U}1}  =  \frac{1}{2} \left(a_{\tilde{U}} + c - \Delta_{\tilde{U}} \right), \qquad
  \lambda_{\tilde{U}2}  =  \frac{1}{2} \left(a_{\tilde{U}} + c + \Delta_{\tilde{U}}\right) \ ,
\end{eqnarray}
where again we have renamed $x = 1,\ 2,\ 3$ by $\tilde{U} = \tilde{u},\ \tilde{c},\ \tilde{t}$, and
$\Delta_{\tilde{U}} = + \sqrt{(a_{\tilde{U}}-c)^2+4b^2}$. The mixing angles for $R_{\tilde{U}}$ become
\begin{eqnarray} \label{anglesb}
  \sin\theta_{\tilde{U}} = \frac{1}{\sqrt{2}}\left( 1+\frac{a_{\tilde{U}}-c}{\Delta_{\tilde{U}}}\right)^{1/2},
  \qquad
  \cos\theta_{\tilde{U}} = \frac{1}{\sqrt{2}}\left( 1-\frac{a_{\tilde{U}}-c}{\Delta_{\tilde{U}}}\right)^{1/2} \ ,
\end{eqnarray}
and so we finally arrive at the desired up-type squark mass matrix
\begin{eqnarray}\label{finalb}
  \tilde{M}^2_{uD} = V^{\dagger} \tilde{M}^2_u V = {\rm diag}
  \left(\lambda_{\tilde{u}1} \ \lambda_{\tilde{u}2} \ \lambda_{\tilde{c}1} \ \lambda_{\tilde{c}2} \ 
  \lambda_{\tilde{t}1} \ \lambda_{\tilde{t}2} \right) \ .
\end{eqnarray}

If we now consider a hierarchy for the
$M_{LL}^2$ eigenvalues of the form $0< a_{\tilde{u}} < a_{\tilde{c}} < a_{\tilde{t}} < c$, then
this implies that $\Delta_{\tilde{u}} > \Delta_{\tilde{c}} > \Delta_{\tilde{t}}$ which then leads to
\begin{eqnarray}\label{spectrumb}
  \lambda_{\tilde{u}2} > \lambda_{\tilde{u}1}, \qquad \lambda_{\tilde{c}2} > \lambda_{\tilde{c}1}, \qquad 
  \lambda_{\tilde{t}2} > \lambda_{\tilde{t}1}, \qquad \lambda_{\tilde{t}1} > \lambda_{\tilde{c}1} > \lambda_{\tilde{u}1} \ .
\end{eqnarray}

In this case the lightest u-type squark has squared mass $\lambda_{\tilde{u}1}$.

\subsection{Case c}\label{casec}
Finally we consider a scenario where both 
$M_{LL}^2$ and $M_{RR}^2$ are general hermitian matrices while $M_{LR}^2$ proportional to the identity ${\cal I}_3$ matrix: 
(again, we show the analysis for the up-type squark mass matrix):
\begin{eqnarray}\label{relationc}
M^2_{LL} = (M^2_{LL})^{\dagger}, \qquad M^2_{RR} = (M^2_{RR})^{\dagger} , \qquad M^2_{LR} = b {\cal I}_3  .
\end{eqnarray}
Then, the mass matrix becomes (see Eq.~(\ref{mu}))
\begin{eqnarray}\label{muc}
  \tilde{M}_u^2 = 
  \left(
  \begin{array}{cc}
    M_{LL}^2 & b {\cal I}_3 \\
    b {\cal I}_3 & M_{RR}^2
  \end{array}
  \right) \ ,
\end{eqnarray}
and using Eq.~(\ref{mstar}) we immediately obtain
\begin{eqnarray}\label{mstarc}
  \tilde{M}^2_{u*}=U^{\dagger} \tilde{M}_u^2 U =   
  \left(
  \begin{array}{cc}
    U_L^{\dagger} M_{LL}^2 U_L & b \ U_L^{\dagger} {\cal I}_3 U_R \\
    (b \ U_L^{\dagger} {\cal I}_3 U_R)^{\dagger} & U_R^{\dagger} M_{RR}^2 U_R
  \end{array}
  \right) = 
  \left(
  \begin{array}{ccc|ccc}
    a_1 & 0 & 0 & b & 0 & 0 \\
    0 & a_2 & 0 & 0 & b & 0 \\
    0 & 0 & a_3 & 0 & 0 & b \\
    \hline
    b & 0 & 0 & k a_1 & 0 & 0 \\
    0 & b & 0 & 0 & k a_2 & 0 \\
    0 & 0 & b & 0 & 0 & k a_3 
  \end{array}
  \right) \ ,
\end{eqnarray}
where in order to obtain the identity in the off-diagonal sub-blocks it is necessary to require
$U_R = U_L$. This in turn requires $M_{LL}^2$ and $M_{RR}^2$ to be diagonalized by the same unitary
matrix and we have chosen the simplest case where they are proportional, i.e. $M_{RR}^2 = k M_{LL}^2$.
Again $a$, $b$, $k \in \Re$ (since all matrices are hermitian).

Applying the method we obtain
\begin{eqnarray} \label{eigenvaluesc}
  \lambda_{\tilde{U}1}  =  \frac{1}{2} \left(a_{\tilde{U}}(k+1) - \Delta_{\tilde{U}}\right), \qquad
  \lambda_{\tilde{U}2}  =  \frac{1}{2} \left(a_{\tilde{x}}(k+1) + \Delta_{\tilde{U}}\right) \ ,
\end{eqnarray}
where  again we have renamed $x = 1,\ 2,\ 3$ by $\tilde{U} = \tilde{u},\ \tilde{c},\ \tilde{t}$, and
$\Delta_{\tilde{U}} = + \sqrt{a_{\tilde{U}}^2(k-1)^2+4b^2}$, and
\begin{eqnarray} \label{anglesc}
  \sin\theta_{\tilde{U}} = \frac{1}{\sqrt{2}}\left( 1+\frac{a_{\tilde{U}}(1-k)}{\Delta_{\tilde{U}}}\right)^{1/2},
  \qquad
  \cos\theta_{\tilde{U}} = \frac{1}{\sqrt{2}}\left( 1-\frac{a_{\tilde{U}}(1-k)}{\Delta_{\tilde{U}}}\right)^{1/2} \ .
\end{eqnarray}
In this case the up-type physical squark mass matrix is given by
\begin{eqnarray}\label{finalc}
  \tilde{M}^2_{uD} = V^{\dagger} \tilde{M}^2_u V = {\rm diag}
  \left(\lambda_{\tilde{u}1} \ \lambda_{\tilde{u}2} \ \lambda_{\tilde{c}1} \ \lambda_{\tilde{c}2} \ 
  \lambda_{\tilde{t}1} \ \lambda_{\tilde{t}2} \right) \ .
\end{eqnarray}

Taking $k>0$ and the 
$M_{LL}^2$ eigenvalues with hierarchy $0< a_{\tilde{u}} < a_{\tilde{c}} < a_{\tilde{t}}$ implies
that $\Delta_{\tilde{u}} < \Delta_{\tilde{c}} < \Delta_{\tilde{t}}$ which then leads to
\begin{eqnarray}\label{spectrumc}
  \lambda_{\tilde{u}2} > \lambda_{\tilde{u}1}, \qquad \lambda_{\tilde{c}2} > \lambda_{\tilde{c}1}, \qquad 
  \lambda_{\tilde{t}2} > \lambda_{\tilde{t}1}, \qquad \lambda_{\tilde{t}2} > \lambda_{\tilde{c}2} > \lambda_{\tilde{u}2} \ .
\end{eqnarray}

We see that in this case the heaviest u-type squark has squared mass $\lambda_{\tilde{t}2}$.

\section{Discussion of FCNC} \label{discussion}
The study of FCNC is crucial to determine the viability of any given parametrization of the sfermions mass matrices\cite{Misiak:1997ei}.
Generally, large off-diagonal terms in the squark mass matrices are strongly constrained by $K^0 - \bar{K}^0$, $D-\bar{D}$,
and $B-\bar{B}$ mixing, as well as by the processes $b\to s\gamma$, $b\to s\bar{l}l$, and $K^0 \to \mu^+\mu^-$ decays (large 
off-diagonal terms in the slepton mass matrix are restricted by $\mu \to e\gamma$). This is one of the reasons why in the
simplified parameter space of the MSSM these off-diagonal terms are simply put to zero. However, it is important to keep
in mind that the actual form of these mass matrices is unknown and that attempts to build them from more general
contexts might in fact lead to more interesting matrices with richer phenomenology. We note that general formulae
exist for the analysis of FCNC processes and they have been obtained either with a general diagonalization or 
through the use of the mass insertion method~\cite{Gabbiani:1988rb,Gabbiani:1996hi}, however, a study considering specific textures for the sfermion 
mass matrices has only been done in~\cite{DiazCruz:2001gf}.

As mentioned above, some interesting modifications to the usual scenarios discussed in the literature can be analyzed easily within the
framework of our method. Take for instance the following assumptions:
\begin{eqnarray} \label{mfv1} 
  m^2_{Qij} & = & m^2_{Qi} \delta_{ij}, \qquad
  m^2_{Uij} = m^2_{Ui} \delta_{ij}, \qquad  m^2_{Dij} = m^2_{Di} \delta_{ij}, \\
  m^2_{Lij} & = & m^2_{Li} \delta_{ij}, \qquad m^2_{Eij} = m^2_{Ei} \delta_{ij},
\end{eqnarray}
and
\begin{eqnarray}\label{mfv2}
  A_{uij} = A_u Y_{uij}, \qquad A_{dij} = A_d Y_{dij}, \qquad  A_{eij} = A_e Y_{eij},
\end{eqnarray}
where $ m^2_{Q}$, $m^2_{U}$, $m^2_{D}$, $m^2_{L}$ and $m^2_{E}$ are such that
$M_{LL}^2$ and $M_{RR}^2$ are proportional to the identity matrix. This
setting corresponds to the case a described in Section~\ref{method} and thus can be
analyzed immediately. The spectrum, for instance, is already given by
Eq.~(\ref{spectruma}). This corresponds to a modification of the Minimal Flavor Violation
(MFV) scenario discussed in~\cite{Chung:2003fi}.

We stress that the method described in this paper can be used to extract general expressions for the main FCNC processes 
listed above in terms of the parametrizations discussed in the previous section. One important observation
is that by analyzing the general expressions for the sfermion mass matrices within this method one can then keep
all the sub-leading terms in the mass matrices and mix the exact diagonalization with the mass insertion method.
We believe this will be helpful in exploring a richer set of extensions and/or
reparametrizations of the MSSM and the work is currently under preparation~\cite{ADN}).

\section{Conclusions} \label{concl}

A method of diagonalization for the sfermion mass matrices has been presented. It assumes hermiticity of the sfermion mass
matrices and works for matrices $M_{LL}^2$, $M_{RR}^2$, and $M_{LR}^2$ such that
\begin{eqnarray}
U_L^{\dagger} M_{LL}^2 U_L = M_{LLD}^2 \qquad U_R^{\dagger} M_{RR}^2 U_R = M_{RRD}^2 \qquad U_L^{\dagger} M_{LR}^2 U_R = M_{LRD}^2 ,
\end{eqnarray}
where $M_{LLD}^2$, $M_{RRD}^2$, and $M_{LRD}^2$ are diagonal matrices. Three specific cases have been presented in the
paper that represent extensions of the constrained MSSM scenario where universality of sfermion masses is assumed and
all the off-diagonal A-terms contributions are set to zero. The spectrum is presented for each case. A model independent
study of FCNC processes is underway and will be presented elsewhere.

\section*{Acknowledgments}
This work was supported in part by CONACYT and SNI. We thank Ricardo S\'aenz for useful discussions. R.N-P. 
acknowledges the \emph{Universidad de Colima} for its warm hospitality and financial support from 
CONACYT through the program \emph{Apoyo Complementario para la Consolidaci\'on Institucional
de Grupos de Investigaci\'on (Retenci\'on)}.


\appendix 
\section{General conditions on the mass matrices}\label{appendix}
The method described in this paper works for hermitian matrices $M_{LL}^2$, $M_{RR}^2$, and $M_{LR}^2$ such that
\begin{eqnarray}\label{relations}
U_L^{\dagger} M_{LL}^2 U_L = M_{LLD}^2 \qquad U_R^{\dagger} M_{RR}^2 U_R = M_{RRD}^2 \qquad U_L^{\dagger} M_{LR}^2 U_R = M_{LRD}^2 ,
\end{eqnarray}
where $M_{LLD}^2$, $M_{RRD}^2$, and $M_{LRD}^2$ are diagonal matrices.

These relations impose the following conditions on the three matrices: let $u_i^L$ denote the eigenvectors
of $M_{LL}^2$ (i.e. the columns of $U_L$) and $u_i^R$ those of $M_{RR}^2$,  then since 
$\langle u_j^{R,L}\ , \ u_k^{R,L} \rangle = \delta_{jk}$, we obtain the desired relations provided
\begin{eqnarray} \label{requirement}
M_{LR}^2 u_i^R = \lambda_i u_i^L, \qquad  M_{LR}^2 u_i^L = \lambda_i u_i^R \ .
\end{eqnarray}

Thus, $M_{LR}^2$ mixes the eigenvectors of matrices $U_L$ and $U_R$. Note that the matrices in Eqs.~(\ref{relationa}),
(\ref{relationb}), and (\ref{relationc}) trivially satisfy these conditions.


\begin{thebibliography}{99}
\bibitem{Haber:1984rc}
  H.~E.~Haber and G.~L.~Kane,
  Phys.\ Rept.\ {\bf 117}, 75 (1985).
  
\bibitem{Aitchison:2005cf}
  I.~J.~R.~Aitchison,
  arXiv:hep-ph/0505105.
  
\bibitem{Chung:2003fi}
  D.~J.~H.~Chung, L.~L.~Everett, G.~L.~Kane, S.~F.~King, J.~D.~Lykken and L.~T.~Wang,
  Phys.\ Rept.\  {\bf 407}, 1 (2005)
  [arXiv:hep-ph/0312378].
  
\bibitem{Misiak:1997ei}
  M.~Misiak, S.~Pokorski and J.~Rosiek,
  Adv.\ Ser.\ Direct.\ High Energy Phys.\ {\bf 15}, 795 (1998)
  [arXiv:hep-ph/9703442].

\bibitem{Kobayashi:2000br}
  T.~Kobayashi and O.~Vives,
  Phys.\ Lett.\  B {\bf 506}, 323 (2001)
  [arXiv:hep-ph/0011200].
  
\bibitem{DiazCruz:2001gf}
  J.~L.~Diaz-Cruz, H.~J.~He and C.~P.~Yuan,
  Phys.\ Lett.\ B {\bf 530}, 179 (2002)
  [arXiv:hep-ph/0103178].
  
\bibitem{GomezBock:2008hz}
  M.~Gomez-Bock,
  Rev.\ Mex.\ Fis.\ {\bf 54}, 30 (2008)
  [arXiv:0810.4309 [hep-ph]].
  
\bibitem{Hiller:2008sv}
  G.~Hiller, Y.~Hochberg and Y.~Nir,
  arXiv:0812.0511 [hep-ph].

\bibitem{Gabbiani:1988rb}
  F.~Gabbiani and A.~Masiero,
  Nucl.\ Phys.\  B {\bf 322}, 235 (1989).

\bibitem{Gabbiani:1996hi}
  F.~Gabbiani, E.~Gabrielli, A.~Masiero and L.~Silvestrini,
  Nucl.\ Phys.\  B {\bf 477}, 321 (1996)
  [arXiv:hep-ph/9604387].
  
\bibitem{ADN} Alfredo Aranda, J.L. D\'iaz-Cruz, R. Noriega-Papaqui, in preparation.
  
  

\end{thebibliography}
\end{document}